\documentstyle[aps,prl,epsf]{revtex}

\begin{document}
\draft
\preprint{IUCM99-003}
\twocolumn[\hsize\textwidth\columnwidth\hsize\csname
@twocolumnfalse\endcsname
\title{Spin Bottlenecks in the Quantum Hall Regime}

\author{A.H. MacDonald}

\address{Department of Physics, Indiana University, Bloomington,
         IN 47405 }

\date{\today}
\maketitle
\begin{abstract}

We present a theory of time-dependent tunneling between a metal and a partially
spin-polarized two-dimensional electron system (2DES). We find that the leakage
current which flows to screen an electric field between the metal and the
2DES is the sum of two exponential contributions whose relative weights
depend on spin-dependent tunneling conductances, on quantum corrections to the
electrostatic capacitance
of the tunnel junction, and on the rate at which the 2DES spin-polarization
approaches equilibrium. For high-mobility and homogeneous 2DES's at Landau level
filling factor $\nu=1$, we predict a ratio of the fast and slow leakage rates
equal to $(2K+1)^2$ where $K$ is the number of reversed spins in the skyrmionic
elementary charged excitations.

\end{abstract}
\pacs{}
\vskip2pc]

There has been renewed\cite{johnson} interest recently in non-equilibrium spin
accumulation\cite{oldspinaccum} due to electronic transport in 
spin-polarized electron systems, in part because these accumulations are
important in giant magnetoresistance\cite{gmr}. 
In this paper we address spin-accumulation
in the tunneling current between a metal and a two-dimensional
electron system (2DES) in the quantum Hall regime. Our work is motivated in part
by recent experiments\cite{ashoori} which have discovered unexplained two-rate
leakage currents for such tunnel junctions, and also in part by the
long\cite{awschalom} spin-relaxation times of 2DES's, especially long in the
quantum Hall regime\cite{vonklitzing,barrett,theory}.  
We find that spin accumulation depends subtly on the interplay of
spin-dependent tunneling conductances, thermodynamic densities-of-states, and
spin relaxation rates.  According to our theory, the double-rate current 
found in experiment signals sizable quantum corrections to the
effective capacitance of the junction {\em and} spin-relaxation
processes which depart from those in conventional metals, both
expected in the quantum Hall regime.
Gapped quantum Hall states lead to rapid
variation of the chemical potential\cite{qhcapexp} with density, and the
presence 
of Skyrmion elementary charged excitations\cite{skyrmrefs} requires an unusual
spin-relaxation process.  We predict that, for homogeneous 2DES and thin
tunneling barriers, the ratio of fast and slow relaxation rates
at $\nu=1$ will equal $(2K+1)^2$ , where $K$ is the number of reversed
spins in the lowest energy Skyrmion quasiparticles.

We start from the following phenomenological linear response equations
which we believe to be broadly applicable for tunneling between a metal and a
2DES.
\begin{eqnarray}
\dot{Q}_{\uparrow} &=& -\mu_{\uparrow}G_{\uparrow} + (\mu_{\downarrow} -
\mu_{\uparrow}) G_{s} \nonumber \\
\dot{Q}_{\downarrow} &=& -\mu_{\downarrow} G_{\downarrow}+
(\mu_{\uparrow} - \mu_{\downarrow}) G_{s}.
\label{phenomenology}
\end{eqnarray}
Here $Q_{\sigma}$ and $\mu_{\sigma}$ are
the spin-$\sigma$ particle-number and chemical potential in the 2DES.
These equations express
spin-partitioning of the tunneling current; the assumption of separate chemical
potentials for the 2DES spin subsystems is valid when the spin-relaxation
time is much longer than other characteristic scattering times in the 2DES. In
these equations we have placed the zero of energy at the chemical potential of
the metal. The two terms on the right hand side of Eqs.~[\ref{phenomenology}]
account respectively for tunneling across the junction and relaxation of the
2DES spin subsystems toward mutual equilibrium. A closed description of electron
transport in the system requires, in addition to the above {\it conductance}
equations, which relate currents to chemical potential differences, a set of
{\it capacitance} equations which relate these chemical potentials to
accumulated charges:
\begin{eqnarray}
\mu_{\uparrow} &=& -V_{0} + (C^{-1})_{\uparrow\uparrow} Q_{\uparrow} +
(C^{-1})_{\uparrow\downarrow} Q_{\downarrow}\\
\mu_{\downarrow} &=& -V_{0} + (C^{-1})_{\downarrow\uparrow} Q_{\uparrow} +
(C^{-1})_{\downarrow\downarrow} Q_{\downarrow}
\end{eqnarray}
Here $V_{0}$ represents the electrostatic contribution from charges external to the
2DES.  Elements of the inverse capacitance $(C^{-1})$ matrix have an electrostatic
contribution proportional to the width of the tunnel barrier and
a quantum `chemical potential' contribution due to the Fermi statistics and
correlations of electrons in the 2DES:
\begin{equation}
(C^{-1})_{\sigma\sigma'} = \frac{1}{C_{g}} + \frac{1}{A}\;
\frac{d {\tilde \mu}_{\sigma}}{dn_{\sigma'}} \equiv \frac{1}{C_{g}} +
F_{\sigma,\sigma'}
\end{equation}
where $C_{g} = A \epsilon / (4 \pi e^{2} d) $ is the electrostatic 
capacitance of the junction, $A$ is the
cross-sectional area of the two-dimensional electron system,
$\epsilon$ is the dielectric constant of the host semiconductor,
$d$ is the distance between the metallic electrode and the 2DES,
and $\tilde \mu_{\sigma}$ is the spin-$\sigma$ chemical
potential of the 2DES relative to its electric subband energy.
The notation above is motivated by the 
relationship between $F_{\sigma,\sigma'}$
and Fermi liquid theory interaction parameters.
In the commonly employed Hartree mean-field approximation,
\begin{equation}
F_{\sigma,\sigma'} =  \frac{\delta_{\sigma,\sigma'}}{A D^{0}_{\sigma}}
\end{equation}
where $D^{0}_{\sigma}$ is the non-interacting 2DES density-of-states per spin.
In general
$d\mu_{\sigma}/d\eta_{\sigma'}$ is non-zero for $\sigma \neq \sigma'$
because of electronic correlations.

\begin{figure}
\epsfxsize=8.5cm \epsffile{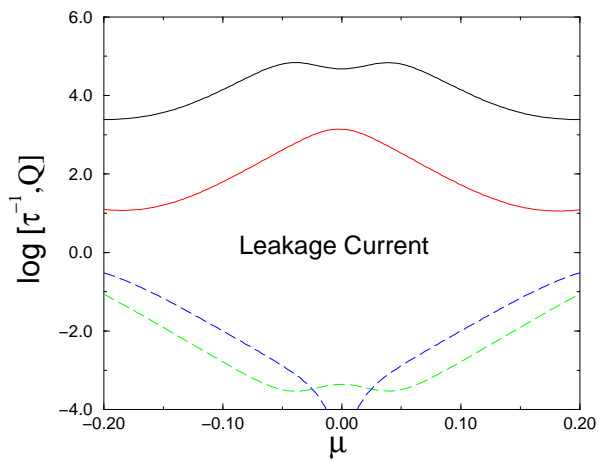} 
\vspace{0.5cm}
\caption[]{Leakage current between a metallic electrode and a 2DES as a function
of chemical potential near $\nu=1$. The chemical potential is in $e^2/\ell$
units, and the zero of energy is chosen so
that$\mu=0$ at $\nu=1$. The solid and dotted lines show leakage rates in
fast and slow channels respectively, while the long and short dashed lines show
the capacitance contributions, $I_+ \tau_+$, and $I_- \tau_-$, from fast and
short rate channels respectively. These curves were calculated using the
non-interacting Skyrmion model explained in the text to evaluate the quantum
inverse capacitance contributions and a separation $d=5 \ell$ between the metal
and the 2DES. $G_{\sigma}$ and $G_s$ were assumed to have the form
$1000/F_{\sigma,\sigma} + 10$ and $1/(F_{\uparrow,\uparrow}
F_{\downarrow,\downarrow}) + 0.01$ in arbitrary units; the two terms
representing uniform system golden rule and inhomogeneity contributions
respectively. The leakage
rates are in units of $C_{g}^{-1}$ times the conductance units and the
capacitances per unit area are in units of $\ell^{-1}$.
This figure is for $k_B T =0.025 e^2/\ell$.}
\label{fig:one}
\end{figure}

These equations can be used to describe various 
time-dependent or ac transport experiments;
we apply them here to the situation studied recently\cite{ashoori}
by Chan {\em et al.} in which a chemical potential difference across
the junction is created by external charges and the leakage current which 
reestablishes equilibrium is measured as a function of time.
In an obvious matrix notation we rewrite the conductance and capacitance
equations as $\dot{\bf Q} = {\bf G}{\bf \mu}$ and 
${\bf \mu} = {-\bf V_{0}} + {\bf C^{-1}} {\bf Q}$.
Eliminating the chemical potentials using the capacitor
equations yields a set of
two coupled first-order inhomogeneous linear differential equations for the
time-dependent spin-up and spin-down charges in the 2DES.
Solving these with the boundary condition
$Q_{\uparrow}(t=0) = Q_{\downarrow}(t=0) =0$ yields for the spin-dependent
currents into the 2DES:
\begin{equation}
\dot{Q}_{\sigma}(t) = I_{\sigma,+} \exp( - t/\tau_{+} ) + I_{\sigma,-}
\exp ( -t/\tau_{-})
\label{ioft}
\end{equation}
where $\tau_{+}^{-1}$ and $\tau_{-}^{-1}$, generalized RC relaxation rates
, are the eigenvalues of ${\bf A} = {\bf G} {\bf C^{-1}}$.
We have obtained the following explicit expressions for
$I_{\uparrow,\pm}$:
\begin{eqnarray}
\frac{I_{\uparrow,+}}{V_{0}} &=& \frac{G_{\uparrow}}{2} \left( 1 +
\frac{A_{\uparrow,\uparrow}-A_{\downarrow,\downarrow}}{\tau_{+}^{-1}
- \tau_{-}^{-1}} \right) + \frac{G_{\downarrow} A_{\uparrow,\downarrow}}{\tau_{+}^{-1}
- \tau_{-}^{-1}} \\
\frac{I_{\uparrow,-}}{V_{0}} &=& \frac{G_{\uparrow}}{2} \left( 1 -
\frac{A_{\uparrow,\uparrow}-A_{\downarrow,\downarrow}}{\tau_{+}^{-1}
- \tau_{-}^{-1}} \right)  - \frac{G_{\downarrow} A_{\uparrow,\downarrow}}{\tau_{+}^{-1}
- \tau_{-}^{-1}}.
\label{weights}
\end{eqnarray}
The corresponding expressions for $I_{\downarrow,\pm}$ are
obtained by interchanging spin labels.  In Fig.~[\ref{fig:one}] 
we show results obtained for the dependence of leakage current on
2DES chemical potential near Landau level filling factor 
$\nu =1$ which follow from a non-interacting Skyrmion model
of the 2DES described below.  These replicate all major 
features found in experiment\cite{ashoori}.
The peak in both fast and slow relaxation rates is due to the 
sharp decrease in capacitance as the $\nu=1$ incompressible
quantum Hall state is approached.  The leakage current is dominated
by the slow process, except in a narrow range very close to 
$\nu=1$ where the fast process takes over.  The origin of
this crossover is explained below.

\begin{figure}
\epsfxsize=8.5cm \epsffile{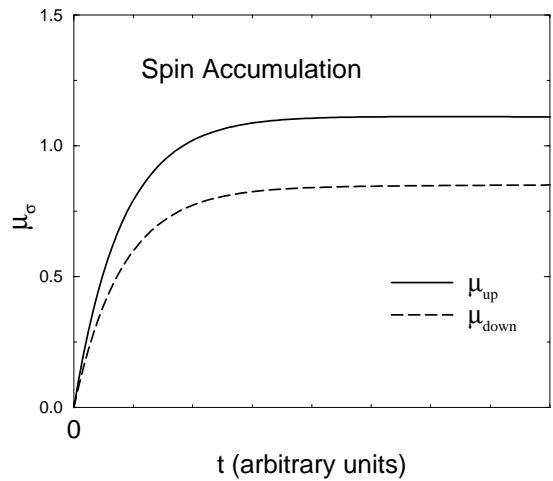} 
\vspace{0.5cm}
\caption[]{
$\mu_{\sigma} +V_0$ in units of $V_0$ as a function of time for
the model of Fig.~[\ref{fig:one}] at $\mu=0.05 e^2/\ell$.  The maximum
time illustrated is, $10^{-4}$ units so that $t/\tau_{+}$ 
large but $t/\tau_{-}$ is still small at the largest times.

}
\label{fig:two}
\end{figure}

Similar results can be obtained for the instantaneous chemical
potentials of the spin-up and spin-down subsystems:
\begin{equation}
\mu_{\sigma}(t) = -V_0 + \sum_{s}\mu_{\sigma,s}(1 - \exp (-t/\tau_{s}))
\label{cpoft}
\end{equation}
where $\mu_{\sigma,\pm} = \sum_{\sigma'} C^{-1}_{\sigma,\sigma'} I_{\sigma',\pm} \tau_{\pm}$.
We note that $\mu_{\sigma,+}+\mu_{\sigma,-}= V_{0}$; current flows until
the electrochemical
potential change for {\it each} spin cancels the electric
potential from external charges.  The two spin subsystems are in
equilibrium at both the beginning and the end of the relaxation
process, but are, in general, out of equilibrium at intermediate
times.  The non-equilibrium spin accumulation 
$\mu_{\uparrow}(t) - \mu_{\downarrow}(t) =
(\mu_{\downarrow,-}-\mu_{\uparrow,-})
(\exp (-t/\tau_{-}) - \exp(-t/\tau_{+}))$. 
What is readily
separated in experiment are the fast and slow relaxation contributions to
the current, not the spin subsystem contributions.  Nevertheless, we see
that non-equilibrium spin accumulations occur system between times
$\tau_{-}$ and $\tau_{+}$ whenever both contributions are present.
In Fig.~[\ref{fig:two}] we plot, time-dependent chemical potentials
calculated for the non-interacting Skrymion model at $\mu =0.05 (e^2/\ell) $ .  The chemical
potentials change quickly on a time scale $\tau^{+}$ and 
$\mu_{\uparrow}$ overshoots $V_0$; on a much longer time scale 
the two chemical potentials approach $V_0$ from opposite sides.
We note from Fig.~[\ref{fig:one}] that the slow leakage current process
actually dominates the capacitance at this value of $\mu$;
the naive view that the fast process is current flow to the
2DES while the slow process is spin-equilibration is incorrect.

Before turning to the quantum Hall regime, where non equilibrium
spin-accumulations
are large, it is instructive to examine several limits for which
spin-accumulation does not occur.  For $F_{\sigma,\sigma'} C_{g} << 1$
we find that $\tau_{+}^{-1} = (G_{\uparrow} +
G_{\downarrow})/C_{g}$, \, $\tau_{-}^{-1} \to 0$, $I_{\sigma,+} = G_{\sigma}
V_{0}$ and $I_{\sigma,-} \to 0$.  In this limit, which usually holds
for metallic electrodes, spin-independent
electrostatic contributions dominate electrochemical potential changes; 
no non-equilibrium spin-accumulation occurs because the spin
subsystems are not driven from equilibrium by electrostatic potentials.
For strong tunnel barriers ($G_{\sigma} << G_{s}$), on the other hand,
$\tau_{+}^{-1} = G_{s} (F_{\uparrow,\uparrow} + F_{\downarrow,\downarrow} -
2F_{\uparrow,\downarrow})$,\, $\tau_{-}^{-1} =
(G_{\uparrow}+G_{\downarrow}) (F_{\uparrow,\uparrow} F_{\downarrow,\downarrow} -
F_{\uparrow,\downarrow}^{2})/(F_{\uparrow,\uparrow}
+ F_{\downarrow,\downarrow} -
2 F_{\uparrow,\downarrow})$ and that the fast relaxation current
$I_{+} \equiv I_{\uparrow,+} + I_{\downarrow,+} =0$.  
For this limit spin accumulation does not occur 
because the relaxation processes are fast enough to maintain
instantaneous equilibrium.  Unlike the electrostatic-dominance case 
discussed first, the leakage current flows at the
slow rate $\tau_{-}^{-1}$.  A third more subtle limit in which
spin-accumulation does not occur applies to  
Fermi gas 2DES's in which we may ignore correlation 
contributions to the chemical potential and the commonly 
adopted forms $G_{\sigma} = c
D_{\sigma}$, \, $G_{s} = c_{s} D_{\uparrow} D_{\downarrow }$
hold.  These expressions result from golden rule estimates
of quasiparticle tunneling and spin-flip transition rates
respectively , $c$ is a constant which declines
exponentially with the thickness of the tunneling barrier and
$c_{s}$ is a constant dependent on spin-orbit scattering strength in
the 2DES.  For this model we find
that $\tau_{+}^{-1} = c (1/C_{g} + D_{\uparrow}
+ D_{\downarrow})$, $\tau_{-}^{-1} = (c + c_{s}(D_{\uparrow} +
D_{\downarrow})/C_{g}$ and all the weight is in the fast leakage 
current.  No spin-accumulation occurs because the
spin subsystems are not coupled by interactions and the ratio of
tunneling conductances equals the ratio of the rates at which
the chemical potentials increase with density.  Finally we
mention the case in which the 2DES is paramagnetic to which we
return below.  For $G_{\uparrow} = G_{\downarrow}$ and $F_{\uparrow,\uparrow}
= F_{\downarrow,\downarrow}$, symmetry forbids
spin accumulations.  An explicit calculation finds no weight for
the slow leakage current and the rate ratio 
\begin{equation}
\frac{\tau_{-}}{\tau_{+}} =
\frac{ G_{\uparrow} (2/C_{g}+F_{\uparrow,\uparrow}+
F_{\uparrow,\downarrow})}{(G_{\uparrow} +
2 G_{s})(F_{\uparrow,\uparrow}-F_{\uparrow,\downarrow})}.
\label{rateratio}
\end{equation}

None of these limits apply throughout the quantum Hall regime.
Near integer Landau level filling factors, Fermi statistics and
correlations in the 2DES,
not electrostatics, dominate the electrochemical potential
changes with density\cite{qhcapexp}.  Equilibrium electronic states
contain\cite{skyrmrefs} complex Skyrmion quasiparticles
whose formation from the fully spin-polarized $\nu=1$ ground state
cannot be achieved by a single-particle process. Spin-equilibration
will therefore be slow\cite{skyrmrefs}.  The two spin systems are
intricately coupled so that the Fermi gas limit does not apply.
Furthermore, the 2DES will generally be strongly
spin-polarized.

\begin{figure}
\epsfxsize=8.5cm \epsffile{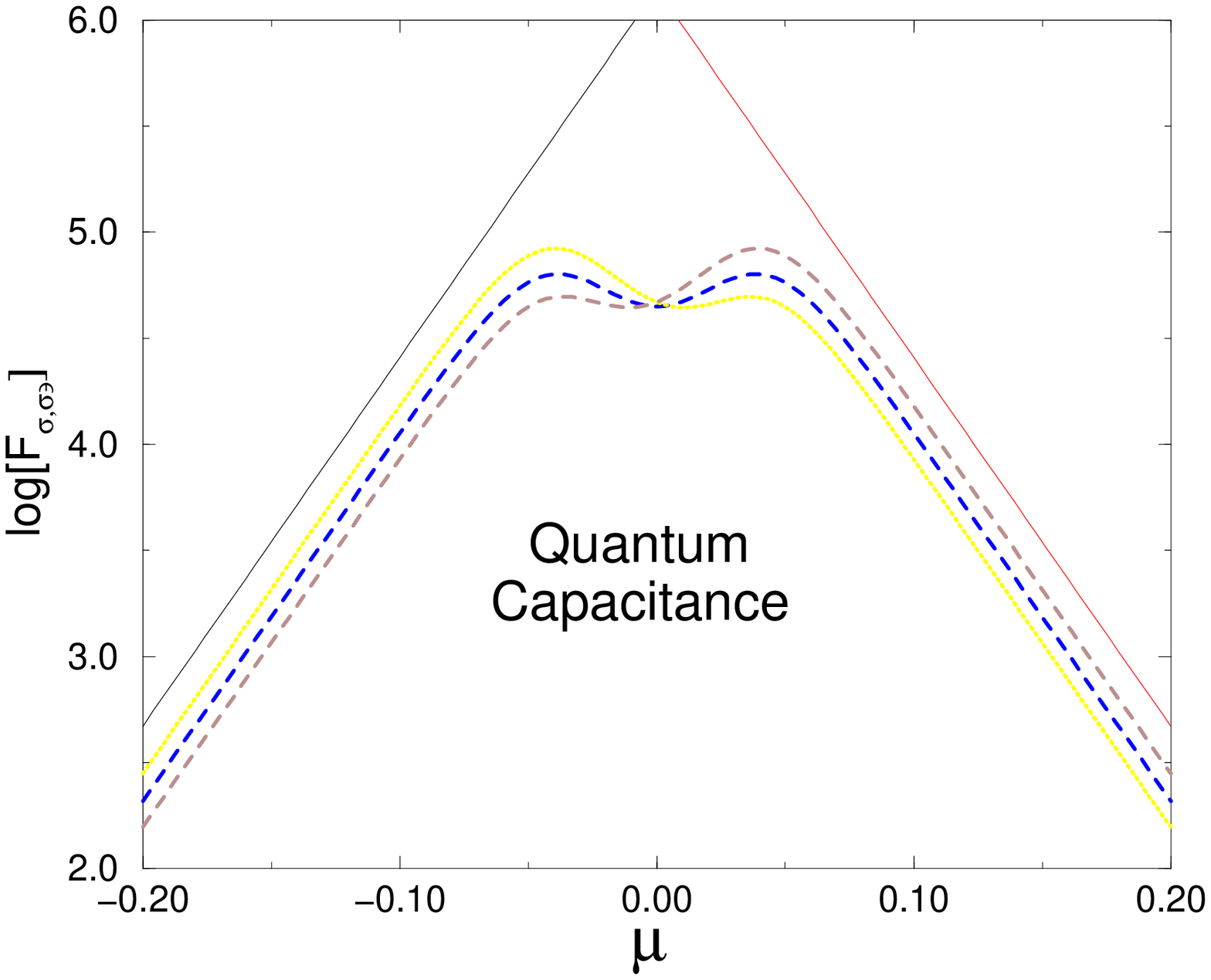} 
\vspace{0.5cm}
\caption[]{
Quantum contributions to the inverse capacitance for a 
2DES near $\nu=1$.  The solid lines show the majority 
and minority spin results for $F_{\sigma,\sigma}$ for 
a non-interacting model which contains only $K=0$ quasiparticles.
For the Skrymion model, $F_{\uparrow,\uparrow}$,
$F_{\downarrow,\downarrow}$ and $F_{\uparrow,\downarrow}$ are
shown by dotted, short-dashed, and long-dashed lines respectively.
}
\label{fig:three}
\end{figure}

A simple model of the 2DES valid at low temperature 
for $\nu$ near one, is obtained by ignoring interactions between
Skyrmions.  We obtain the following grand-canonical ensemble
expressions for the occupation
probabilities of the $N_{\phi} = A/ (2 \pi \ell^{2})$ Skyrmion
quasielectron and quasihole states with $K$ excess reversed spins:
\begin{eqnarray}
n_{Ke} &=& f(\epsilon_{K} + K \mu_{\uparrow} - (K+1)
\mu_{\downarrow} ) \\
n_{Kh} &=& f(\epsilon_{K} + (K+1) \mu_{\uparrow} - K
\mu_{\downarrow}).
\label{skyrmionoccs}
\end{eqnarray}
Here $f(\epsilon) = (\exp(\epsilon/k_{B}T)+1)^{-1}$ is a Fermi
factor\cite{caveat2}, $\epsilon_{K}$ is the energy of a Skyrmion
quasiparticle, $(2 \pi  \ell^{2})^{-1} $ is the density of a full Landau
level, and we have chosen the zero of energy so that
quasielectron and quasihole skyrmion states have the same
energy\cite{kunyang}. When the spin subsystems are in equilibrium
($ \mu_{\uparrow} = \mu_{\downarrow} = \mu $) we can
use Eqs.~(\ref{skyrmionoccs}) to calculate the chemical potential, given the
Landau level filling factor.  Eqs.~(\ref{skyrmionoccs}) express
the property that the $K$-th quasielectron Skyrmion 
is formed by adding $K+1$ spin-down electrons and removing $K$
spin-up electrons from to the $\nu=1$ ground state.
For non-interacting electrons only $K=0$ quasiparticles occur;
for typical 2DES's, on the other hand, the lowest energy
quasiparticles have $K=3$\cite{palacios}. From
Eqs.~(\ref{skyrmionoccs}) we calculate the following
thermodynamic densities of states, $D_{\sigma,\sigma'} \equiv
d n_{\sigma} / d \mu_{\sigma'} $ for the coupled spin systems:
\begin{eqnarray}
2 \pi \ell^{2} D_{\uparrow,\uparrow} &=& 
\sum_{K} \big[ (K+1)^{2} \Delta(\epsilon_{K} + \mu) +K^{2}
\Delta(\epsilon_{K} -\mu) \big] \nonumber \\
 2 \pi \ell^{2} D_{\uparrow,\downarrow} &=& - 
\sum_{K} K (K+1) \left[ \Delta(\epsilon_{K}+\mu) +
\Delta(\epsilon_{K} - \mu) \right] \nonumber 
\end{eqnarray}
where $\Delta(x) = {\rm sech}^{2}(x/2k_B T)/4 k_{B} T$.
In Fig.~[\ref{fig:three}] we plot quantum contributions to
the model's inverse capacitance,
obtained by inverting this density-of-states matrix.
The large peaks in capacitance near $\nu=1$ are responsible for the peaks in 
both leakage current rates.  In the low-temperature limit,
only the lowest energy $K=3$ skyrmion will contribute so that 
for $\nu > 1$, $F_{\uparrow,\uparrow}$ \, $F_{\downarrow,\downarrow}$,
and $F_{\uparrow,\downarrow}$ occur in the ratio 
$(K+1)^2:K^2:K(K+1)$ = $16:9:12$.  This contrasts with the
non-interacting electron case for which $F_{\uparrow,\uparrow}$ is 
much larger than $F_{\downarrow,\downarrow}$ and
$F_{\uparrow,\downarrow}$ vanishes.  (The same ratios apply for
$\nu < 1$ with inverted spin-indices.)  For $\nu =1$ the 
low-temperatures equilibrium state charge is added to the $\nu=1$
state in the form of $K=3$ skyrmions, {\it i.e.} for each 4 up-spins
added to the 2DES, three down spins are removed.  It follows that 
the time-integrated spin-up and spin-down leakage
currents are approximately equal in magnitude and opposite in sign.

In contrast to the partitioning of total leakage charge between spins
, which is determined purely by thermodynamic considerations, 
the observable partitioning between fast and slow components
is difficult to understand intuitively in the general case.  
Simplification occurs, however, when $\nu \equiv 1$.
It follows from particle-hole symmetry\cite{kunyang},
that the symmetries for leakage into a paramagnetic 2DES
also hold at $\nu=1$, explaining the vanishing weight 
in the slow leakage current channel seen in Fig.~[\ref{fig:one}] 
as the filling factor approaches one. 
The ratio of fast to slow leakage rates, (Eq.~(\ref{rateratio})
depends only on thermodynamic quantities provided that the tunneling
barrier is thin enough that $G_{\sigma} >> G_{s}$. 
Then, provided that the temperature is sufficiently low that 
quantum terms dominate the inverse capacitance we find that 
$\tau_{-}/\tau_{+} = (2K+1)^2 = 49$, in rough agreement with the 
findings of Chan {\em et al.}\cite{ashoori}.   We ascribe discrepancies
to the inhomogeneity present at integer filling factors in 
all current samples. 

I thank Ho Bun Chan and Ray Ashoori for stimulating discussions and 
the ITP at UC Santa Barbara for hospitality during a Quantum Hall
Workshop.  This work was supported by NSF Grants DMR-9714055 and PHY94-07194.

\begin{references}

\bibitem{johnson} M. Johnson and R.H. Silsbee, Phys. Rev.
Lett. {\bf 55}, 1790 (1985); M. Johnson, Phys. Rev. Lett.
{\bf 70}, 2142 (1993).

\bibitem{oldspinaccum} A. G. Aronov, Pis'ma Zh. Eksp. Teor. Fiz.
{\bf 24}, 37 (1976) [JETP Lett. {\bf 24}, 32 (1976)];
P.M. Tedrow and R. Merservey, Phys. Rep.
{\bf 238}, 174 (1994).

\bibitem{gmr} T. Valet, and A. Fert, Phys. Rev. B {\bf 48}, 7099 (1993);
M.A.M. Gijs and G.E.W. Bauer, Adv. Phys. {\bf 46}, 285 (1997);
J-Ph Ansermet, J. Phys. C {\bf 10}, 6027 (1998).

\bibitem{ashoori} H.B. Chan, R.C. Ashoori, L.N. Pfeiffer, 
and K.W. West, preprint (1999).

\bibitem{awschalom} J.M. Kikkawa and D.D. Awschalom,
Phys. Rev. Lett. {\bf 80}, 4313 (1998).

\bibitem{vonklitzing} A. Berg, M. Dobers, R.R. Gerhardts, and K. v. Klitzing,
Phys. Rev. Lett. {\bf 64}, 2563 (1990); S. Kronmuller, W. Dietsche, 
K. v. Klitzing, G. Denninger, W. Wegscheider, and M. Bichler,
Phys. Rev. Lett. {\bf 82}, 4070 (1999).

\bibitem{barrett} N.N. Kuzma, P. Khandelwal, S.E. Barrett, L.N. Pfeiffer,
and K.W. West, Science {\bf 281}, 686 (1998).

\bibitem{theory} I.D. Vagner and T. Maniv,
Phys. Rev. Lett. {\bf 61}, 1400 (1988);
Dimitri Antoniou and A.H. MacDonald, Phys. Rev. B {\bf 43}, 11686 (1991).

\bibitem{qhcapexp} T.P. Smith, W.I. Wang, and P.J. Stiles, Phys. Rev. B
{\bf 34}, 2995 (1986); V. Mosser, D. Weiss, K. v. Klitzing, K. Ploog, and
G. Weimann, Solid State Commun. {\bf 58}, 5 (1986);
S.V. Kravchenko, D.A. Rinberg, S.G. Semenchinsky, and V.M. Pudalov,
Phys. Rev. B {\bf 42}, 3741 (1990); J.P. Eisenstein, L.N. Pfeifer,
and K.W. West, Phys. Rev. Lett. {\bf 68}, 674 (1992).
 
\bibitem{skyrmrefs} S.L. Sondhi, A. Karlhede, S.A. Kivelson, and
E.H. Rezayi, Phys. Rev. B {\bf 47}, 16419 (1993); H.A. Fertig, L. Brey,
R. C\^ot\'e, and A.H. MacDonald, Phys. Rev. B {\bf 50}, 11018 (1994);
H.A. Fertig, L. Brey, R. C\^ot\'e, and A.H. MacDonald,
Phys. Rev. Lett. {\bf 77}, 1572 (1996).

\bibitem{caveat2} We use Fermi statistics for the Skrymions 
to crudely represent their repulsive interactions.  These calculations
are all in the dilute Skyrmion limit where the actual particle statistics
play at most a minor role.

\bibitem{kunyang} Kun Yang and A.H. MacDonald, Phys. Rev. B
{\bf 51}, 17247 (1995).

\bibitem{palacios} J.J. Palacios, D. Yoshioka, and A.H. MacDonald, 
Phys. Rev. B {\bf 54}, R2296 (1996).  

\end{references}
\end{document}